\documentclass[11pt,paper]{JHEP3}
\usepackage{epsfig}
\usepackage{latexsym}
\usepackage{amssymb}
\usepackage{amsmath}
\usepackage{graphicx}
\usepackage{dcolumn}
\usepackage{bm}

\renewcommand{\vec}[1]{\boldsymbol{#1}}
\newcommand{\be}{\begin{equation}}
\newcommand{\ee}{\end{equation}}
\newcommand{\ba}{\begin{eqnarray}}
\newcommand{\ea}{\end{eqnarray}}
\newcommand{\bi}{\begin{itemize}}
\newcommand{\ei}{\end{itemize}}
\newcommand{\tr}{{\rm Tr\,}}
\newcommand{\re}{\mathop{\rm Re}}

\newcommand{\<}{\langle}
\renewcommand{\>}{\rangle}
\newcommand{\eq}{Eq.~}

\newcommand{\fig}{Fig.~}

\newcommand{\la}{\label}

\newcommand{\txts}{\textstyle}

\newcommand{\im}{\mathop{\rm Im}}

\newcommand{\bp}{\boldsymbol{p}}

\newcommand{\bz}{\boldsymbol{z}}

\title{Implications of Poincar\'e symmetry\\ for thermal field theories in finite-volume}

\author{Leonardo Giusti \\
Dipartimento di Fisica, Universit\`a di Milano--Bicocca,\\
and INFN, sezione di Milano--Bicocca,\\
I-20126 Milano, Italy\\
\email{Leonardo.Giusti@mib.infn.it}}

\author{Harvey B. Meyer \\
PRISMA Cluster of Excellence, \\ 
Institut f\"ur Kernphysik and Helmholtz Institute Mainz\\
Johannes Gutenberg-Universit\"at Mainz\\
D-55099 Mainz, Germany\\
\email{meyerh@kph.uni-mainz.de}}

\abstract{
The analytic continuation to an imaginary velocity $i\vec\xi$
of the canonical partition function of a thermal system
expressed in a moving frame has a natural implementation
in the Euclidean path-integral formulation in terms of
shifted boundary conditions. 
Writing the Boltzmann factor as
$\exp[-L_0(\widehat H-i\vec\xi\cdot\widehat {\vec P})]$,
the Poincar\'e invariance  underlying a relativistic theory 
implies a dependence of the free-energy 
on $L_0$ and the shift $\vec\xi$ 
only through the combination $\beta= L_0 \sqrt{1+\vec \xi^2}$. 
This in turn implies a set of Ward identities, some of which
were previously derived by us, among the correlators of the 
energy-momentum tensor. 
In the infinite-volume limit they lead to relations among
the cumulants of the total energy distribution and those 
of the momentum, i.e.\
they connect the energy and the momentum distributions 
in the canonical ensemble.
In finite volume the Poincar\'e symmetry
translates into exact relations among partition functions 
and correlation functions
defined with different sets of (generalized) 
periodic boundary conditions.
They have interesting applications in lattice field theory. 
In particular, they offer Ward identities to 
renormalize non-perturbatively the energy-momentum tensor
and novel ways to compute thermodynamic potentials.
At fixed bare parameters they also provide a simple
method to vary the temperature in much smaller steps 
than with the standard procedure.}

\begin{document}

\section{Introduction}
It is a recurring theme in quantum field theory that a 
symmetry has far reaching consequences even when it is 
softly broken\footnote{Here with ``softly'' we refer 
to any breaking which does not modify the renormalization 
pattern of the theory, e.g.\ mass terms and 
(generalized) periodic boundary conditions.}. In this 
paper we show that, for a relativistic theory set up on a 
space with one or more dimensions of finite length and 
(generalized) periodic boundary conditions\footnote{Translational 
invariance is thus preserved.}, the underlying Lorentz 
symmetry leads to interesting consequences. 

When continued analytically to an imaginary velocity 
vector $\vec v=i \vec\xi$, where $\vec \xi \in\mathbb{R}^3$, 
the canonical partition function of a thermal field theory
formulated in a moving frame has a straightforward
definition in the functional integral formalism. It is 
the ordinary Euclidean path integral with shifted boundary 
conditions in the time-direction~\cite{Giusti:2010bb,Giusti:2011kt}.
In the zero-temperature limit and in presence of a mass gap, 
the invariance of the theory (and of its vacuum) under the 
Poincar\'e group forces its free-energy to be independent of the shift 
$\vec \xi$. At non-zero temperature the finite time-length $L_0$ 
breaks the euclideanized Lorentz group softly, consequently the free energy 
depends on the shift (velocity) explicitly but only 
through the combination $\beta= L_0 \sqrt{1+\vec \xi^2}$. 
An interesting set of Ward identities (WIs) 
follows. As shown in section \ref{sec:InfVol}, they provide
a recursion relation among the cumulants of the momentum 
distribution~\cite{Giusti:2010bb,Giusti:2011kt}, they relate the 
total energy and momentum distributions in the rest-frame, and 
they suggest new ways to compute thermodynamic potentials.
These results generalize those found in Ref.~\cite{Giusti:2011kt} 
to a generic theory and to a generic value of the shift
$\vec\xi$. For the clarity of the presentation and 
to avoid unessential technical complications, however, we restrict 
ourselves to bosonic theories in this paper.

The considerations above extend to a thermal theory set up 
in a finite spatial volume with periodic boundary conditions.
The independence of the free energy on the angle between 
the time and the space directions is replaced by relations 
among partition functions of systems with the time and 
the spatial extensions $L_k$ Lorentz transformed, e.g. 
\be\la{eq:Zequiv}
Z(L_0,L_1,L_2,L_3;\xi_1) = 
Z(\frac{L_1}{\sqrt{1+\xi_1^2}},L_0 \sqrt{1+\xi_1^2},L_2,L_3;-\xi_1)
\ee
with $\vec \xi=(\xi_1,0,0)$. The WIs are readily 
extended to finite-volume systems.

These properties find interesting applications when a 
theory is discretized on the lattice, where a non-zero 
shift can easily be implemented~\cite{Giusti:2010bb}.
Lorentz invariance, which is recovered in the continuum limit only,   
allows one to vary the temperature of the system by 
changing either $\vec \xi$ or $L_0$, i.e.  in much \emph{smaller} steps 
(at fixed bare parameters) with respect to varying $L_0$ alone.
Thanks to the misalignment of the lattice axes with respect to 
the periodic directions, the WIs provide new ways to compute 
thermodynamic potentials numerically and new conditions 
to renormalize non-perturbatively the energy-momentum tensor.

\section{Thermal field theory in a moving frame\la{sec:InfVol}}
In this section we focus on properties of a relativistic thermal 
system in the infinite-volume limit. In a moving frame, the 
total energy and momentum densities are given by (\cite{LandauLif6}, paragraph 133)
\be\la{eq:EPtransf}
e' = \frac{1}{1-\vec v^2}\,(e+\vec v^2 p)\; ,\qquad 
\vec p' = \frac{e+p}{1-\vec v^2} \, \,\vec v\, ,
\ee
where $\vec v$ is the velocity of the center-of-mass relative to the
observer, $e$ and $p$ are the energy density and pressure in 
the rest frame respectively. The enthalpy density $(e+p)$ in the
rest frame plays the role of the inertial mass density of the 
system, and its rest volume appears contracted by a factor 
$\sqrt{1-\vec v^2}$ in the moving frame. The standard definition 
of the partition function is\footnote{We use the notation $L_0$
because this parameter represents the length of the 
Euclidean time direction in the path integral formalism.} (\cite{LandauLif10}, paragraph 2)
\be\la{eq:calZ}
{\cal Z}(L_0,\vec v) \equiv \tr\{ e^{-L_0\, (\widehat H-\vec v\cdot \vec {\widehat P})}\}
\; , 
\ee
where $\widehat H$ and $\vec {\widehat P}$ are the Hamiltonian and the total 
momentum operator expressed in a moving frame. We focus on 
the Euclidean formulation, where it is natural to continue ${\cal Z}$
to imaginary velocities $\vec v=i \vec\xi$ with the Lorentz group 
 replaced by SO(4). The partition function  
\be\la{eq:Z}
Z(L_0,\vec\xi) = \tr\{e^{-L_0(\widehat H-i\vec\xi\cdot\widehat {\vec P})}\} 
\ee
corresponds to the ordinary Euclidean path integral with 
shifted boundary conditions in the 
time-direction~\cite{Giusti:2010bb,Giusti:2011kt}. The free-energy density 
can be defined as usual
\be\label{eq:freeE} 
f(L_0,\vec \xi) = - \frac{1}{L_0 V} 
\ln Z(L_0,\vec \xi)\; ,
\ee
where $V$ is the volume observed in the moving frame. In 
the thermodynamic limit the invariance of the dynamics 
under the SO(4) group implies
\be\la{eq:fEucl.eq.fcan}
f(L_0,\vec \xi) = f(L_0\sqrt{1+\vec \xi^2},\vec 0)\; .
\ee
In section~\ref{sec:tftsbc} we derive this equation in 
the path integral formalism starting from a finite-volume system, 
and provide the functional form for the finite-volume corrections.
\eq(\ref{eq:fEucl.eq.fcan}) is consistent with modern thermodynamic
arguments on the Lorentz transformation of the temperature and the
free energy ~\cite{Ott,Arz} (the issue has been debated
for a long time, see~\cite{Prz} for a recent discussion).
Before entering into the details of the derivation it is interesting to
discuss the origin of this formula, and to anticipate some of the  
implications of the rich kinematics in the boundary 
conditions that Lorentz symmetry allows for in the path-integral 
formulation.

\subsection{Ward identities for the total energy and momentum}
Relation (\ref{eq:fEucl.eq.fcan}) is the source of certain WIs for 
the energy-momentum tensor, some of which were already derived 
in Ref.~\cite{Giusti:2010bb,Giusti:2011kt}. They can be generated in 
a quasi-automated fashion by deriving the free-energy density with 
respect to $L_0$ and $\xi_k$. By remembering that 
the cumulants of the total momentum distribution can be 
written as~\cite{Giusti:2010bb}
\be\label{eq:cum1}
k_{\{2 n_1, 2 n_2, 2 n_3\}} \!\equiv\! \frac{1}{V}\langle {\widehat P}_1^{2 n_1} 
{\widehat P}_2^{2 n_2} {\widehat P}_3^{2 n_3} \rangle_c \! 
=\! \frac{(-1)^{n_1+n_2+n_3+1}}{L_0^{2 n_1+2 n_2+2 n_3-1}}
\frac{\partial^{2n_1}}
{\partial \xi_1^{2n_1}} \frac{\partial^{2n_2}}{\partial \xi_2^{2n_2}}
\frac{\partial^{2n_3}}{\partial \xi_3^{2n_3}}
f(L_0,\vec \xi)\Big|_{\vec \xi=0},\!\!\!\!
\ee
in the thermodynamic limit a plethora of Ward identities among on-shell 
correlators of the total momentum and/or energy are derived by inserting 
\eq(\ref{eq:fEucl.eq.fcan}) in (\ref{eq:cum1}). By choosing 
$\vec \xi=\{\xi_1,0,0\}$, it is straightforward to derive the master 
equation
\be\label{eq:fnd1}
\frac{k_{\{2 n, 0, 0\}}}{L_0} =(-1)^{n+1}\, (2n-1)!!\,
\Big\{\frac{1}{L_0}\frac{\partial}{\partial L_0} \Big\}^n f(L_0,\vec \xi)\Big|_{\vec \xi=0}  
\qquad n=1,2,\dots \; . 
\ee
If we remember that in the Euclidean the momentum operator maps to  
$\widehat P_k \rightarrow -i {\overline T}_{0k}$, where 
${\overline T}_{\mu\nu}(x_0) = \int d^3 x\, T_{\mu\nu}(x)$ with 
$T_{\mu\nu}$ being the energy-momentum field of the theory,
an immediate consequence of \eq(\ref{eq:fnd1}) is the recursion 
relation ($x_0^i$ all different) \
\be
\langle {\overline T}_{01}(x_0^1) 
\dots {\overline T}_{01}(x_0^{2n}) \rangle_c =  
(2 n -1)\, \frac{\partial}{\partial L_0}\,\Big\{\frac{1}{L_0}\,
\langle {\overline T}_{01}(x_0^3) 
\dots {\overline T}_{01}(x_0^{2n}) \rangle_c \Big\}\quad n=2,3\dots\; 
\ee
which extends to a generic theory the derivation presented 
for the scalar one in Ref.~\cite{Giusti:2011kt}. 
Cumulants with non-null indices 
in the other directions can be related to those in 
(\ref{eq:fnd1}) by cubic symmetry.\\[0.175cm]
\indent 
If we define $c_1 \equiv e-f$ and recall that the higher cumulants of the 
total energy distribution are given by
\be\la{eq:cn}
c_n \equiv \frac{1}{V}\, \langle\, \widehat H^n\, \rangle_c = (-1)^{n+1}\left[ n\, \frac{\partial^{n-1}}{\partial L_0^{n-1}}
+ L_0\, \frac{\partial^n}{\partial L_0^n} \right]
f(L_0,\vec \xi)\Big|_{\vec \xi=0} \quad n=2,3\dots\; ,
\ee
it is clear that there is a linear relation among 
$c_1,\dots, c_n$ and the $n$ first derivatives of 
the free-energy density. 
Since \eq(\ref{eq:fnd1}) gives the $k_{\{2 n, 0, 0\}}$ as 
linear combinations of the very same derivatives, a linear relation exists 
among the $n$ first non-trivial cumulants of the energy and momentum 
distributions in the thermodynamic limit. 
Some details of the required combinatorics are summarized in 
appendix \ref{sec:s2apdx}. The 
result reads\footnote{Notice that the coefficients multiplying $c_\ell$ 
are all positive.} 
\be\la{eq:k2ncn}
k_{\{2n,0,0\}} = \frac{(2n-1)!!}{(2 L_0^2)^n}\,\sum_{\ell=1}^n 
\frac{(2n-\ell)!}{\ell!(n-\ell)!}\, {(2 L_0)^\ell\, c_\ell }\, ,
\ee
and it shows that the total energy and momentum distributions 
of a relativistic thermal theory are related. Up to $n=4$ 
we obtain
\ba\label{eq:nice}
L_0\, k_{\{2,0,0\}\,} &=& c_1\;,\nonumber\\
L_0^3\, k_{\{4,0,0\}}\, &=& 9\, c_1 + 3\,L_0\,  c_2\;,\\
L_0^5\, k_{\{6,0,0\}} &=& 225\, c_1 + 90\,L_0\,  c_2 + 
15\, L_0^2\, c_3\;,\nonumber\\
L_0^7\, k_{\{8,0,0\}} &=& 11025\, c_1 + 4725\, L_0\,  c_2 + 
1050\,L_0^2\,  c_3 + 105\,L_0^3\,  c_4\;.\nonumber
\ea
As expected $c_1=e+p$ is necessarily positive. 
Since $L_0^2\,  c_2$ is the specific heat, the fourth cumulant of the momentum 
turns out to always be positive. If we remember that in the Euclidean
$\langle T_{00} \rangle = -e$ and $\langle T_{kk} \rangle = p$, 
Eqs.~(\ref{eq:nice}) can also be written as 
\ba\label{eq:contWIs}
L_0\, \langle {\overline T}_{01}\, T_{01}\rangle_c 
& = & \langle T_{00} \rangle - \langle T_{11} \rangle \; ,\nonumber\\[0.125cm]
L_0^3\, \langle {\overline T}_{01}\, {\overline T}_{01}\, 
                {\overline T}_{01}\, T_{01}\rangle_c & = &
9\, \langle T_{11} \rangle - 9\, \langle T_{00} \rangle 
+ 3\, L_0\,  \langle {\overline T}_{00} T_{00} \rangle_c\; ,\\
& \dots & \nonumber
\ea 
where in each correlator the energy-momentum fields are inserted  
at different times. These WIs generalize to all cumulants of a 
generic field theory those found in 
Refs.~\cite{Giusti:2010bb,Giusti:2011kt}. They show that the 
thermodynamics of a relativistic
theory can be studied from its thermal momentum 
distribution and vice-versa.

\subsection{Ward identities in presence of a non-zero shift}
When $\vec \xi \neq 0$ parity is softly broken by the boundary 
conditions in the compact direction, odd derivatives in the $\xi_k$ 
do not vanish anymore, and new interesting WIs hold. By deriving 
once with respect to $L_0$ and $\xi_k$, it is easy to 
obtain the first non-trivial relation
\be\label{eq:WIodd}
\langle T_{0k} \rangle_{\vec\xi} = \frac{\xi_k}{1-\xi_k^2} 
\left\{\langle T_{00} \rangle_{\vec\xi}  
- \langle T_{kk} \rangle_{\vec \xi}\,\right\}\; .
\ee
An interesting consequence of this equation is that the 
entropy density $s$ of the system at the inverse temperature 
$\beta=L_0\sqrt{1+{\vec \xi}^2}$ is given by
\be\label{eq:spv}
s= - \frac{L_0}{\gamma^3\xi_k}\, \langle T_{0k} \rangle_{\vec\xi}
\ee 
which, by following Refs.~\cite{Giusti:2010bb,Giusti:2011kt}, 
can also be written as 
\be
s = - \frac{1}{V\gamma^3\xi_k}
\frac{\partial}{\partial\xi_k} \ln Z(L_0,\vec\xi)\; 
\ee
where $\gamma=1/\sqrt{1+\vec\xi^2}$. Ward identities among correlators 
with more fields can easily be 
obtained by considering higher order derivatives in $L_0$ and $\xi_k$.
For instance by deriving two times with respect to the shift components, by using 
\be
L_0 \langle \; {\overline T}_{0k}(L_0) \, O \rangle_{\vec\xi,\, c}
= \frac{\partial}{\partial \xi_k} \< O \>_{\vec\xi} 
\ee
where $O$ is a generic field with support located at a physical distance 
from the time-slice $L_0$, we obtain
\be\label{eq:wipv}
\langle T_{0k}\rangle_{\vec\xi}  = 
\frac{L_0 \xi_k}{2}\,  \sum_{ij}\,
\left\langle {\overline T}_{0i}\,  T_{0j} \right\rangle_{\vec\xi,\, c}\, 
\left[ \delta_{ij} - \frac{\xi_i\, \xi_j}{\vec\xi^2}\right]\; . 
\ee
This equation and \eq(\ref{eq:WIodd}) can be enforced in regularizations 
that break translational invariance, such as the lattice, to renormalize 
non-perturbatively the traceless components of the energy-momentum 
tensor, see section \ref{sec:appl}. By combining 
Eqs.~(\ref{eq:spv}) and (\ref{eq:wipv}),
the entropy density can also be computed as 
\be\label{eq:stwopt}
\displaystyle
s^{-1} =  -  
\frac{\gamma^3}{2}\,  \sum_{ij}\,
\frac{\left\langle{\overline T}_{0i}\,  T_{0j} \right\rangle_{\vec\xi,\, c}}
{\langle T_{0i} \rangle_{\vec\xi} \langle T_{0j}\rangle_{\vec\xi}}\, \xi_i \xi_j 
\,\Big[ \delta_{ij} - \frac{\xi_i \xi_j}{\vec \xi^2}\Big]\; 
,
\ee
and the analogous expression for the specific heat reads 
\be\label{eq:cvtwopt}
\displaystyle
\frac{c_v}{s^2} =  -  
\frac{\gamma^3}{2}\,  \sum_{ij}\,
\frac{\left\langle{\overline T}_{0i}\,  T_{0j} \right\rangle_{\vec\xi,\, c}}
{\langle T_{0i} \rangle_{\vec\xi} 
\langle T_{0j}\rangle_{\vec\xi}}\, \frac{\xi_i \xi_j}{\vec\xi^2} 
\,\Big[ (1-2\vec\xi^2)\delta_{ij} - 3 \frac{\xi_i \xi_j}{\vec \xi^2}\Big]\; .
\ee

\bigskip
\section{Generalized periodic boundary conditions\la{sec:GPBC}}
Consider a quantum field theory\footnote{Since the considerations 
in this section are valid for a generic number of dimensions $d>1$, 
we leave the value of $d$ unspecified. In the rest of the 
paper the results of this section will be used for $d=4$.} 
defined on $\mathbb{R}^d$, an orthonormal basis, 
and $d$ linearly independent 
\emph{primitive vectors} $ v^{(\mu)}$ ($\mu=0,1,\dots,d-1$).  
The latter can be represented by a 
\emph{primitive matrix} $V\in {\rm GL}(d,\mathbb{R})$ 
whose columns are the components of $v^{(\mu)}$ in the orthonormal basis.  
For a given point labeled with the coordinates $x_\mu$, the
field is identified at all points with coordinates
\be
x_\mu + V_{\mu{\nu}} m_{\nu},\qquad m_\nu\in\mathbb{Z}\; ,
\ee
i.e. we impose generalized periodic boundary conditions (GPBCs)
\footnote{For brevity we refer to a theory satisfying GPBCs 
as a finite-volume theory.}. The shifted boundary conditions
which implement the partition function in \eq(\ref{eq:Z}) are 
a special case of GPBCs. 

In addition to the parameters already present in infinite volume, 
the finite-volume theory contains $d^2$ extra parameters specifying 
the coordinates of the primitive vectors. By defining the primitive 
cell as usual
\be
\Omega = \Big\{ x \in \mathbb{R}^d\,|\;  
       x_\mu = V_{\mu{\nu}} t_{\nu},\;
            0\leq t_\mu  < 1 \Big\}\; ,
\ee
$d(d-1)/2$ parameters specify the orientation of the cell 
relative to the orthonormal basis (in $d=3$ and for
orthogonal primitive vectors, these are the Euler angles), while 
$d(d+1)/2$ fix its geometry, namely the length of the vectors 
$v^{(\mu)}$ and the $d(d-1)/2$ angles between them.
For a Lorentz-invariant theory, the
absolute orientation of the primitive cell is clearly of no
consequence. The partition function of the finite-volume theory 
is unchanged if $V$ is replaced by   
\be\label{eq:Lambda}
 V \rightarrow \Lambda V, \qquad \Lambda \in {\rm SO}(d)\; , 
\ee
i.e. the $d(d-1)/2$ parameters that specify the orientation of
the cell are redundant. This is the invariance which allows one 
to generalize \eq(\ref{eq:fEucl.eq.fcan}) in finite volume, and 
to derive the corresponding WIs. 
At variance with the infinite-volume case, the partition function of 
a finite-volume theory is also left unchanged under the discrete group 
of transformations SL$(d,\mathbb{Z})$. As is well known from 
crystallography, two geometrically different primitive cells may in fact 
describe the same crystal: any set of vectors $v^{(\mu)}$ that generates 
the same discrete set of points where the fields are identified is 
equivalent. This amounts to the freedom of replacing the matrix $V$ by 
a new matrix whose columns are linear combinations with integer 
coefficients of the columns of $V$, with the restriction that the inverse 
relation exists and contains only integer coefficients.  The latter 
condition requires the determinant of the two primitive matrices to 
be equal up to a sign. Here we restrict ourselves to the transformations with 
positive sign, which maintain the orientation of the unit cell. In short, 
the transformation is
\be\la{eq:SLZtrafo}
V \rightarrow V M,\qquad M\in {\rm SL}(d,\mathbb{Z})\; ,
\ee
i.e. a discrete equivalence between two GPBCs.
To summarize, the most general relation between two primitive matrices
$V$ and $W$ corresponding to a relativistic field theory with 
two different sets of GPBCs and equal partition functions, is given by 
\be\la{eq:equiv}
 W = \Lambda\, V M, \qquad \Lambda\in{\rm SO}(d),
\quad M\in {\rm SL}(d,\mathbb{Z})\; .
\ee
The matrix $M$ modifies the geometry of the primitive cell, while
$\Lambda$ modifies its orientation. The freedom to choose the former
is a property of periodic boundary conditions, the freedom to choose
the latter is a property of the SO($d$) invariance of the
infinite-volume field theory. The relation (\ref{eq:equiv}) defines an
equivalence relation (in the mathematical sense) between the 
primitive matrices $V$ and $W$. We will write the relation $V\sim W$.
In Appendix~\ref{app:MOM} we verify in momentum space that two
partition functions defined by path integrals with a common
Lagrangian density and with equivalent sets of boundary conditions are
equal. More precisely, the actions in the two theories are related by
\be
 S(V;[\phi]) = S(\Lambda V M;[\phi^\Lambda]), 
\ee
where $[\phi^\Lambda]$ means that every field of the theory has been rotated.
Correlation functions of fields can also be mapped between equivalent 
descriptions of the same system by taking into account their transformation 
properties under the SO($d$) group. It is also interesting to notice 
that, by an appropriate field transformation, the effect of the 
non-orthogonality of the original primitive vectors can be re-absorbed into 
a re-definition of the action, see again appendix~\ref{app:MOM}.

\section{Finite-volume theory with shifted boundary conditions\la{sec:tftsbc}}
The finite-volume analogue of the partition 
function\footnote{In finite volume we will use the primitive matrix as argument of 
the partition function $Z$ and of the free-energy $f$.} (\ref{eq:Z}) 
\be\la{eq:defZEucl}
Z(V_{\rm sbc}) = 
\tr\{e^{-L_0(\widehat H-i\vec\xi\cdot\widehat{\vec P})}\}\; ,
\ee
where
\be
\la{eq:Vsbc}
V_{\rm sbc} = \left(\begin{array}{c@{~~}c@{~~}c@{~~}c}
L_0 &  0  & 0 & 0  \\
L_0\xi_1  & L_1 & 0 & 0 \\
L_0\xi_2  & 0 & L_2 & 0 \\
L_0 \xi_3 & 0 & 0 & L_3 
\end{array}   \right)\; , 
\ee
can be expressed as a 
Euclidean path integral with the fields satisfying 
standard periodic boundary conditions in the 
spatial directions, and shifted boundary conditions~\cite{Giusti:2010bb,DellaMorte:2010yp}
in time\footnote{Relative to these references, we adopt here
a different sign convention for the shift in the path integral.}
\be
\phi(L_0, \vec x) = \phi(0, \vec x - L_0\, \vec \xi)\; .  
\ee
Due to the spatial periodicity, 
$\xi_k'=\xi_k+L_k/L_0$ is equivalent to $\xi_k$, and therefore
the imaginary velocity components can be restricted to the interval
\be
-\frac{L_k}{2L_0}< \xi_k \leq \frac{L_k}{2L_0}\; . 
\ee
For later use it is useful to note the effect of taking derivatives of 
the partition function with respect to the external parameters,
\ba
\la{eq:dV00}
 \<T_{00}\>_{V_{\rm sbc}} &=& \frac{1}{L_0 L_1L_2L_3} \bigg(
L_0\frac{\partial}{\partial L_0}  - \sum_k \xi_k \frac{\partial}{\partial \xi_{k}}\bigg) \ln Z(V_{\rm sbc})\; \nonumber
\\
\<T_{0k}\>_{V_{\rm sbc}} &=& \frac{1}{L_0L_1L_2L_3} \frac{\partial}{\partial \xi_{k}} \ln Z(V_{\rm sbc})
\qquad\qquad\qquad\qquad\quad k=1,2,3
\\
\<T_{kk}\>_{V_{\rm sbc}} &=& \frac{1}{L_0L_1L_2L_3} 
\left(L_k\frac{\partial}{\partial L_k} + \xi_k\frac{\partial}{\partial\xi_k}\right) \ln Z(V_{\rm sbc})
 \qquad k=1,2,3\; ,\nonumber
\ea
and to introduce the notation
\be\label{eq:gamms}
\gamma = \big(1+\vec\xi^2\big)^{-{1}/{2}}, \qquad \quad \gamma_{kl} = \big(1+\xi_k^2+\xi_l^2\big)^{-{1}/{2}},
\qquad \quad \gamma_k = \big(1+\xi_k^2\big)^{-{1}/{2}}\; . 
\ee
By defining 
\be
V_1 = M^{-1} R\, V_{\rm sbc} M = \left(\begin{array}{c@{~~~}c@{~~~}c@{~~~}c}
L_1\gamma_1 & 0 & 0 & 0 \\
-L_1\gamma_1\xi_1 & L_0/\gamma_1 & 0 & 0 \\
0 & L_0\xi_2 & L_2 & 0 \\
0 & L_0\xi_3 & 0 & L_3 \\
\end{array} \right)
\ee
with
\be
R=\left(\begin{array}{c@{~~~}c@{~~~}c@{~~~}c}
\gamma_1 & \gamma_1 \xi_1 & 0 & 0 \\
-\gamma_1 \xi_1 & \gamma_1 & 0 & 0 \\
0 & 0 & 1 & 0 \\
0 & 0 & 0 & 1 \\
\end{array}   \right),
\qquad
M=\left(\begin{array}{c@{~~~}c@{~~~}c@{~~~}c}
0 & 1 & 0 & 0 \\
-1 & 0  & 0 & 0 \\
0 & 0 & 1 & 0 \\
0 & 0 & 0 & 1 \\
\end{array}   \right),
\ee
we conclude from section~\ref{sec:GPBC} that $Z(V_{\rm sbc}) = Z(V_1)$.
We first focus on the case $\xi_2=\xi_3=0$, and later use the 
SO(3) rotation symmetry to generalize the result to a generic shift
vector. The partition function can be interpreted in terms of the 
states that propagate in the direction given 
by the first column of $V_1$.  In the thermal field theory language, the 
latter are the eigenstates of the `screening' Hamiltonian 
$\widetilde H$, which acts on states living on a slice of dimensions 
$(L_0/\gamma_1)\times L_2 \times L_3$
with ordinary periodic boundary conditions. Their spectrum yields the
spatial correlation lengths of the thermal system at inverse
temperature $(L_0/\gamma_1)$. The partition function can thus 
be written as 
\be\la{eq:Zscreen}
Z(V_1) = 
\tr\Big\{\exp{-L_1\gamma_1(\widetilde H + i {\xi_1}\widetilde\omega)}\Big\},
\ee
where $\widetilde\omega$ is the momentum operator along the primitive vector of
length $(L_0/\gamma_1)$.  Its eigenvalues are the Matsubara frequencies
$\omega_n = \gamma_1 \frac{2 \pi n}{L_0}$, $n\in\mathbb{Z}$.  Assuming that
the Hamiltonian $\widetilde H$ has a translationally invariant vacuum
and a mass gap, the right-hand side of \eq(\ref{eq:Zscreen})
becomes insensitive to the phase in the limit $L_1\to\infty$ at fixed
$\xi_1$ (with exponentially small corrections, see below). This in turn 
implies that the free energy densities associated with $V_{\rm sbc}$ and 
$\mbox{diag}(L_1\gamma_1, L_0/\gamma_1,L_2 ,L_3)$ are equal. 
Thanks to the invariance of the infinite-volume theory under 
three-dimensional rotations, this result extends to a generic imaginary 
velocity $\vec\xi$. In the thermodynamic limit the net effect
of the generic shift $\vec\xi$ is thus to lower the temperature from $1/L_0$ to 
$1/\beta=1/(L_0\sqrt{1+{\vec\xi}^2})$, i.e.\ we have 
proved \eq(\ref{eq:fEucl.eq.fcan}). As anticipated in section 
\ref{sec:InfVol}, when the primitive cell dimensions $L_k$ are all 
asymptotically large the system is characterized by a single 
`short' periodic direction of length $\beta = {L_0}\sqrt{1+\vec\xi^2}$
which is interpreted as being its inverse temperature. Its orientation 
is unusual in that it is not aligned along the time-direction, but 
due to its SO(4) symmetry this is irrelevant. 

In a finite-volume the length of the box dimensions are further sources 
of SO(4) soft breakings, and the above analysis is significantly 
more involved. To go straight to the point, let us assume 
again that only $\xi_1\neq 0$. The Euclidean finite-volume counterpart 
of the textbook relations (\ref{eq:EPtransf}) read
\ba
\<T_{00}\>_{V_{\rm sbc}} & = & \gamma_1^2 \Big(\<T_{00}\>_{RV_{\rm sbc}} +
\xi^2_1 \<T_{11}\>_{RV_{\rm sbc}}\Big) 
- 2\xi_1\gamma_1^2 \<T_{01}\>_{RV_{\rm sbc}}\; , \nonumber\\[0.25cm]
\<T_{01}\>_{V_{\rm sbc}} & = & \gamma_1^2 \Big(\<T_{00}\>_{RV_{\rm sbc}}
-\;\<T_{11}\>_{RV_{\rm sbc}}\;\Big) \xi_1 + 
\gamma_1^2 (1-\xi_1^2) \<T_{01}\>_{RV_{\rm sbc}}\; ,
\ea
where in this case 
\be
R V_{\rm sbc}\Big|_{\xi_2=\xi_3=0} = \left(\begin{array}{c@{~~~}c@{~~~}c@{~~~}c}
L_0/\gamma_1     & L_1\gamma_1\xi_1 & 0 & 0 \\
0               & L_1\gamma_1      & 0 & 0 \\
0 & 0 & L_2 & 0 \\
0 & 0 & 0 & L_3 \\
\end{array} \right)\; . 
\ee
In general the term on the r.h.s $\<T_{01}\>_{RV_{\rm sbc}}$ does not 
vanish in finite volume, while it does in the thermodynamic limit where
Eqs.~(\ref{eq:EPtransf}) are reproduced. Thus if we consider a thermal 
system satisfying ordinary spatial periodic boundary conditions `moving' 
at imaginary velocity $\vec\xi$, an attempt to `boost' it back to the 
rest frame modifies its spatial boundary conditions in such a way that 
the momentum density does not vanish. This effect becomes irrelevant 
when the spatial correlation length is finite and the volume becomes 
large (see below). However, a remarkable
property of periodic boundary conditions is that there are discrete
values of the (imaginary) velocity for which the system at rest still obeys
ordinary spatial periodic boundary conditions, and the textbook 
relations in Eqs.~(\ref{eq:EPtransf}) hold. The term $\<T_{01}\>_{RV_{\rm sbc}}$
does vanish \emph{in finite volume} when the 
the parameters of the system satisfy
\be\la{eq:tune}
\frac{L_1\gamma_1^2\xi_1 }{L_0} = q \in \mathbb{Z}\; .
\ee
Indeed \eq(\ref{eq:tune}) implies that 
$\left.R V_{\rm sbc}\right|_{\xi_2=\xi_3=0}\sim {\rm diag}(L_0/\gamma_1, L_1\gamma_1,L_2 ,L_3)$ by an
${\rm SL}(4,\mathbb{Z})$ transformation, see \eq(\ref{eq:SLZtrafo}).

\subsection{Finite-size effects}
It is interesting to ask about the magnitude of finite-size corrections 
to \eq(\ref{eq:fEucl.eq.fcan}). For a generic shift the effect of 
the finite value of $L_1$ in the free energy $f(V_1)$ can be 
quantified as
\be\la{eq:equivFV}
f(V_1) = \frac{1}{L_0L_1L_2L_3}
\Big[ L_1\gamma_1 E_{\rm vac}(\overline V_1)
-\ln \Big(1 + \nu\!\!\!\!\!\!\!\!\!\! \sum_{\textrm{1-particle states} }
e^{-L_1\gamma_1 (E + i \xi_1 p_1)}\Big)\Big]+\dots
\ee
where $\overline V_1$ is the $(0,0)$ minor of the matrix $V_1$
which describes the space on which the eigenstates of the Hamiltonian
$\widetilde H$ are defined. The vacuum energy $E_{\rm vac}(\overline V_1)$ 
on the space $\overline V_1$ corresponds to the free energy of the system 
in the limit $L_1\to\infty$. As indicated in \eq(\ref{eq:equivFV}), 
the leading correction 
\be
{\cal I}_1 = -\frac{\nu}{L_0L_1L_2L_3}\! \sum_{\textrm{1-particle states} }
e^{-L_1\gamma_1 (E + i \xi_1 p_1)}
\ee
comes from one-particle states, where the factor $\nu$ stands for the 
multiplicity of the lightest screening state of mass $M$. The 
allowed momenta in the periodic box described by 
$\overline V_1$ are given by
\be
\vec p = \left(\begin{array}{c} 
\frac{2\pi\gamma_1n_1}{L_0} - \gamma_1 \xi_2 p_2 - \gamma_1 \xi_3 p_3 \\ 
p_2 \\
p_3 \\
 \end{array}\right),\qquad p_2=2\pi \frac{n_2}{L_2},~~~ p_3=2\pi \frac{n_3}{L_3},
\qquad \vec n\in\mathbb{Z}^3.
\ee
In the following we assume that a momentum of order $1/L_0$ always
costs a substantial gap in energy, and therefore set $n_1=0$. Then 
$\vec p$ is orthogonal to the short periodic direction 
$\vec u^\intercal \equiv \left(L_0/\gamma_1, L_0\xi_2, L_0\xi_3 \right)$.
When the box lengths $L_2$ and $L_3$ are large, we expect the 
dispersion relation
of the one-particle states in momenta orthogonal to $\vec u$ 
to be the ordinary relativistic dispersion relation, due to the 
emerging SO(3) rotation symmetry in the space orthogonal to 
$\vec u$. The leading contribution thus reads
\be
{\cal I}_1 = \frac{-\nu}{L_0L_1L_2L_3}\sum_{p_2,p_3} e^{-L_1\gamma_1( \sqrt{M^2+p_2^2+p_3^2+\gamma_1^2(p_2\xi_2+p_3\xi_3)^2}
- i\xi_1\gamma_1 (p_2\xi_2+p_3\xi_3))}\; .
\ee
Using the Poisson summation formula, diagonalizing the quadratic form
under the square root and appropriately rescaling the momentum
integration variables, we arrive at
\be
\la{eq:this}
{\cal I}_1 =  \frac{-\gamma\nu}{L_0L_1\gamma_1} \sum_{m_2,m_3\in\mathbb{Z}} 
\int_{\mathbb{R}^2}\frac{d^2\textbf{\textsf{p}}}{(2\pi)^2}\; 
e^{-L_1\gamma_1\sqrt{M^2+ \textbf{\textsf{p}}^2}+i\textbf{\textsf{p}} \cdot \textbf{\textsf{x}}}
\ee
with 
\be\la{eq:x} 
\textbf{\textsf{x}}^\intercal = \left(\frac{\gamma}{\gamma_1}\frac{
(\xi_2^2+\xi_3^2)L_1\gamma_1^2\xi_1 
-m_2L_2\xi_2 - m_3L_3\xi_3}{\sqrt{\xi_2^2+\xi_3^2}},
\frac{m_2L_2\xi_3 - m_3L_3\xi_2}{\sqrt{\xi_2^2+\xi_3^2}} \right).
\ee
Using the observation that 
\be
\int \frac{d^2\textbf{\textsf{p}}}{(2\pi)^2}
e^{-|x_1| \sqrt{M^2+\textbf{\textsf{p}}^2}+ i \textbf{\textsf{p}}\cdot \textbf{\textsf{x}}} 
= -2\,\frac{|x_1|}{r}\,{\partial_r} \Delta^3(r,M^2)\; , 
\qquad r=(x_1^2+\textbf{\textsf{x}}^2)^{1/2}\; , 
\ee
where $\Delta^3(r,M^2)$ is the propagator of a free massive scalar particle
on $\mathbb{R}^3$, we finally obtain 
\be
{\cal I}_1 = 
\frac{\gamma\nu}{2\pi L_0}
\sum_{m_2,m_3\in\mathbb{Z}} \frac{1}{r} \frac{d}{dr}
\Big[\frac{e^{-Mr}}{r}\Big]_{r=\sqrt{(Q\vec\mu,Q\vec\mu)}}\; ,
\la{eq:resultEmr}
\ee
where $\vec\mu = (L_1,m_2 L_2,m_3 L_3)$, and 
$Q_{ij}=(\delta_{ij} + (\gamma-1)\xi_i\xi_j/\vec\xi^2)$
defines a positive norm which takes into account the Euclidean version of
relativistic length contraction in 
direction $\vec\xi$. The leading contribution to 
${\cal I}_1$ is thus given by 
the value of $(m_2,m_3)$ that minimizes the norm of vector $\textbf{\textsf{x}}$ 
in \eq(\ref{eq:x}), i.e.\footnote{If $L_1\gg L_2,L_3$, several values of $m_2$ and $m_3$
make comparable contributions, and the original representation in terms of a sum over discrete momenta
is more useful. However, in that case the finite-volume effects associated with the 
other directions will be the dominant ones anyhow.}\ $m_2=m_3=0$. The leading finite-volume effect 
associated with $L_1$ is thus 
\be\la{eq:expo}
{\cal I}_1 = -\frac{\nu}{2\pi L_0 L_1^3}\frac{\gamma^3_{23}}{\gamma^2}
\Big[1+m L_1 \frac{\gamma}{\gamma_{23}} \Big]e^{-M L_1 {\gamma/\gamma_{23}}}\; .
\ee
To obtain the finite volume corrections to $E_{\rm vac}(\overline V_1)$
due to the finiteness of the other spatial directions we can proceed 
iteratively as follows. A helpful observation is that in the 
limit $L_1\to\infty$, the shift $-L_1 \gamma_1 \xi_1$ in 
$V_1$ can be ignored if the ground state is translationally invariant.
We thus obtain
\be\la{eq:Vprime}
\gamma_1 E_{\rm vac}(\overline V_1) 
= - \lim_{L_1\to\infty}  \frac{1}{L_1} \ln Z(V'_{\rm sbc}),
\ee
where $V'_{\rm sbc}$ is obtained from $V_{\rm sbc}$ by making the 
two-step substitutions: first treat $\gamma_1$ as independent of 
$\xi_1$ and set
\be
L_0\to L_0/\gamma_1\;, \quad L_1\to L_1\gamma_1\;, \quad 
\xi_1\to 0\;, \quad \xi_2 \to \gamma_1\xi_2\; , 
\quad \xi_3 \to\gamma_1\xi_3\; ,
\ee
and then assign to $\gamma_1$ its value in \eq(\ref{eq:gamms}). 
The partition function in \eq(\ref{eq:Vprime}) can now be interpreted
in terms of states living on slices of constant coordinate $x_2$.
The associated leading finite-size corrections are then given by ${\cal I}_2$ 
which, as expected, matches what one would obtain by cyclically 
permuting the three directions in expression 
(\ref{eq:expo}). In summary, the
leading finite-size contributions to the free energy are
\be\label{eq:finalFV}
f(V_{\rm sbc})-f(L_0\sqrt{1+\vec \xi^2}) = {\cal I}_1 + {\cal I}_2 + 
{\cal I}_3 + \cdots 
\ee
and the larger contribution(s) among those on the r.h.s.
depends on the particular geometry of the shifted boundary conditions. 
As a test of this result, we perform an independent calculation for 
a free-boson theory in appendix \ref{sec:apdxfree}. It
should be noted that, since the leading correction arises from
one-particle states, the leading finite-size effects are predicted 
exactly by a free-boson theory if one sets its mass to $M$. 
The procedure followed in this section and the 
formula (\ref{eq:finalFV}) generalize to shifted boundary conditions 
those in Ref.~\cite{Meyer:2009kn}. Also in this case the leading 
finite-volume corrections are fully determined once the mass $M$ and the 
multiplicity $\nu$ of the lightest screening multiplet are 
known.

\subsection{Ward identities for total energy and momentum\la{sec:WI}}
The equality $Z(V_{\rm sbc})=Z(V_k)$, where $V_k$ is defined analogously 
to $V_1$ for the $k$-direction, can be used to generate WIs for correlators 
of the energy and momentum fields in a quasi-automated fashion.  
It suffices to take derivatives with respect to the parameters of the 
primitive vectors. If we derive once with respect to $\xi_k$, the first 
WI is given by
\be\la{eq:oddFV}
\langle T_{0k} \rangle_{V_{\rm sbc}} +
\frac{1+\xi_k^2}{1-\xi_k^2}\, \langle T_{0k} \rangle_{V_k} = 
\frac{\xi_k}{1-\xi_k^2} \left\{\langle T_{00} \rangle_{V_{\rm sbc}}  
- \langle T_{kk} \rangle_{V_{\rm sbc}}\,\right\}\; .
\ee
The second term on the l.h.s proportional to 
$\langle T_{0k} \rangle_{V_k}$ vanishes in the limit 
$L_k\rightarrow\infty$, and as expected it vanishes also at  
finite $L_k$ if the condition analogous to \eq(\ref{eq:tune})
is satisfied, i.e. $\frac{L_k\gamma_k^2\xi_k }{L_0} = q \in \mathbb{Z}$ . 
By differentiating twice with respect to $\xi_k$ and by setting $\xi_k=0$ we obtain
\be\label{eq:bellissima}
L_0\,\langle {\overline T}_{0k}\, T_{0k}\rangle\rangle_{V_{\rm sbc}, c} - 
L_k\,\langle {\widetilde T}_{0k}\, T_{0k}\rangle\rangle_{V_{\rm sbc}, c} = 
\langle T_{00}\rangle_{V_{\rm sbc}} -\langle T_{kk}\rangle_{V_{\rm sbc}} \; , 
\ee
where all insertions in the same correlator are at a physical
distance from each other and 
\be
{\widetilde T}_{\mu\nu}(w_k) =
\int \Big[\prod_{\rho\neq k} d w_\rho\Big] \, T_{\mu\nu}(w)\; . 
\ee
Analogously the fourth derivative leads to 
\ba  \label{eq:bellissima4}\displaystyle
&& L_0^3\,  \< {\overline T}_{0k} {\overline T}_{0k} {\overline T}_{0k} T_{0k}\>\rangle_{V_{\rm sbc}, c}
- L_k^3\, \< {\widetilde T}_{0k} {\widetilde T}_{0k}
{\widetilde T}_{0k} T_{0k}\>\rangle_{V_{\rm sbc}, c} 
=3\, \Big\{\<T_{00}\>_{V_{\rm sbc}}  - \<T_{kk}\>_{V_{\rm sbc}}\Big\}   +
\\[0.25cm] 
&& 3\, \Big\{L_k\, \<{\widetilde T}_{kk} T_{kk}\>_{V_{\rm sbc}, c} 
- L_0\, \<{\overline T}_{00} T_{00}\>_{V_{\rm sbc}, c} \Big\}
+6\,  \Big\{L^2_0\, \<{\overline T}_{0k} {\overline T}_{0k} T_{00}\>_{V_{\rm sbc}, c}  
- L^2_k\, 
\<{\widetilde T}_{0k} {\widetilde T}_{0k} T_{kk}\>_{V_{\rm sbc}, c} \Big\}\; 
\nonumber
\ea
after some rearrangements of the various terms, having set 
$\xi_k=0$ again and having inserted all fields at a physical distance. 
This derivation extends 
to a generic thermal-field theory results previously obtained in the 
scalar field theory, Eqs. (5.3) and (6.15) in Ref.~\cite{Giusti:2011kt}.
Again, due to the breaking of Lorentz symmetry, 
Eqs.~(\ref{eq:bellissima}) and (\ref{eq:bellissima4}) differ from 
those in (\ref{eq:contWIs}) by terms which vanish in the 
thermodynamic limit.

\section{Applications on the lattice\la{sec:appl}}
The shifted boundary conditions discussed so far
provide an interesting formulation to study
thermal field theories on the lattice. There are 
many applications that can potentially benefit from 
them. In this section we sketch a few examples 
with the computation of thermodynamic potentials in mind.

\subsection{Renormalization of the energy-momentum tensor}
In the continuum, the charges associated with translational symmetries, i.e. 
the total energy and momentum fields, do not need any
ultraviolet renormalization thanks to the Ward identities 
that they satisfy, for a recent discussion see 
Ref.~\cite{Giusti:2011kt} and references therein. On the lattice, 
however, translational invariance is broken down to a discrete 
group and the standard charge discretizations acquire 
finite ultraviolet renormalizations. The renormalization pattern 
of the energy-momentum tensor depends on the theory under 
consideration, since its field content
determines what operators $T_{\mu\nu}$ can mix with. For definiteness 
the discussion below focuses on the SU($N$) Yang--Mills theory, 
but it applies to the scalar field theory as well. 

The energy-momentum field $T_{\mu\nu}$ is a symmetric 
rank-two tensor. Its traceless part is an irreducible representation 
of the SO(4) group. On the lattice, however, the 
diagonal and off-diagonal components of this multiplet  
belong to different irreducible representations of the hypercubic lattice 
symmetry group and therefore renormalize in a different way.
In SU($N$) Yang--Mills theory, they both renormalize multiplicatively.
The WIs (\ref{eq:oddFV}) and (\ref{eq:bellissima}) provide two relations 
among the expectation values of the diagonal and off-diagonal components of the 
energy-momentum tensor. They can be enforced on the lattice 
to compute the overall renormalization constant $Z_T$ of the 
multiplet, and the relative normalization $z_{_T}$ between the 
off-diagonal and the diagonal 
components~\cite{Caracciolo:1989pt,Caracciolo:1988hc},
\be
T_{01}^{\rm R} = Z_{_T} T_{01},\qquad \qquad  
T_{00}^{\rm R}-T_{11}^{\rm R} = Z_{_T} \,z_{_T} \,(T_{00}-T_{11}), \qquad {\rm etc.}
\ee
where the fields with a superscript `R' are the renormalized ones.
There are many ways to implement this strategy in practice. 
A possible choice is to require a primitive matrix
\be\displaystyle
V_T = \left(\begin{array}{c@{~~}c@{~~}c@{~~}c}
L_0 &  0  & 0 & 0  \\
\frac{L_0}{2} & \frac{5}{2} L_0 & 0 & 0 \\
0  & 0 & L & 0 \\
0  & 0 & 0 & L 
\end{array}   \right)\; 
\ee
for which the condition (\ref{eq:tune}) holds, and 
compute $z_{_T}$ as
\be
z_{_T} = \frac{3}{2} 
\frac{\langle T_{01} \rangle_{V_T}}
{\langle T_{00} \rangle_{V_T}  - \langle T_{11} \rangle_{V_T}}\; ,
\ee
while $Z_T$ can be determined from ($x_0\neq y_0$, $x_2\neq y_2$)
\be\label{eq:bellissima5}
\frac{Z_T}{z_{_T}}=
\frac{\langle T_{00}\rangle_{V_T} -\langle T_{22}\rangle_{V_T}}
{L_0\,\langle {\overline T}_{02}(x_0)\, T_{02}(y)\rangle\rangle_{V_T, c} - 
L\,\langle {\widetilde T}_{02}(x_2)\, T_{02}(y)\rangle\rangle_{V_T, c}} 
\; .
\ee
Being fixed by WIs, the finite renormalization constants 
$Z_T$ and $z_{_T}$ depend on the bare coupling constant only.
Up to discretization effects, they are independent of the kinematics 
used to impose them, e.g. the volume, the temperature, the shift 
parameter, $x_0$ etc. Ultimately which WIs and/or kinematics 
yield the most accurate results must be investigated numerically.

\subsection{Calculation of the entropy and specific heat}
Once the relevant renormalization constants are determined, 
the entropy density can be computed from the 
expectation value of $T_{0k}$ on a lattice with shifted boundary 
conditions,
\be\label{eq:rotThMuNu}
s = - \frac{Z_{T} L_0 (1+{\vec \xi}^2)^{3/2}}{\xi_k}\, 
\langle T_{0k} \rangle_{\vec\xi}\;, \qquad \xi_k \neq 0\; ,
\ee
by performing simulations at a single inverse temperature value 
$\beta=L_0\sqrt{1 + \vec\xi^2}$, and at a volume large 
enough for finite-size effects to be negligible. The latter are 
exponentially small in $(M L)$, where $M$ is the lightest 
screening mass of the theory. To properly assess discretization
effects, a set of full-fledged simulations needs to be performed 
at several lattice spacings. A rough idea on their magnitude, however, 
can be obtained in the non-interacting limit of the theory. For the   
SU($N$) Yang--Mills theory discretized with the 
Wilson action and for the `clover' form of the lattice field 
strength tensor \cite{Luscher:1996sc}, discretization effects turn out 
to be rather small, see \fig\ref{fig:T0k}  for the choice 
corresponding to 
$\vec \xi = (1,0,0)$. The details of the calculation are given 
in appendix \ref{sec:<T0k>}.
\FIGURE[t!]{
\centerline{\includegraphics[width=12.0 cm,angle=0]{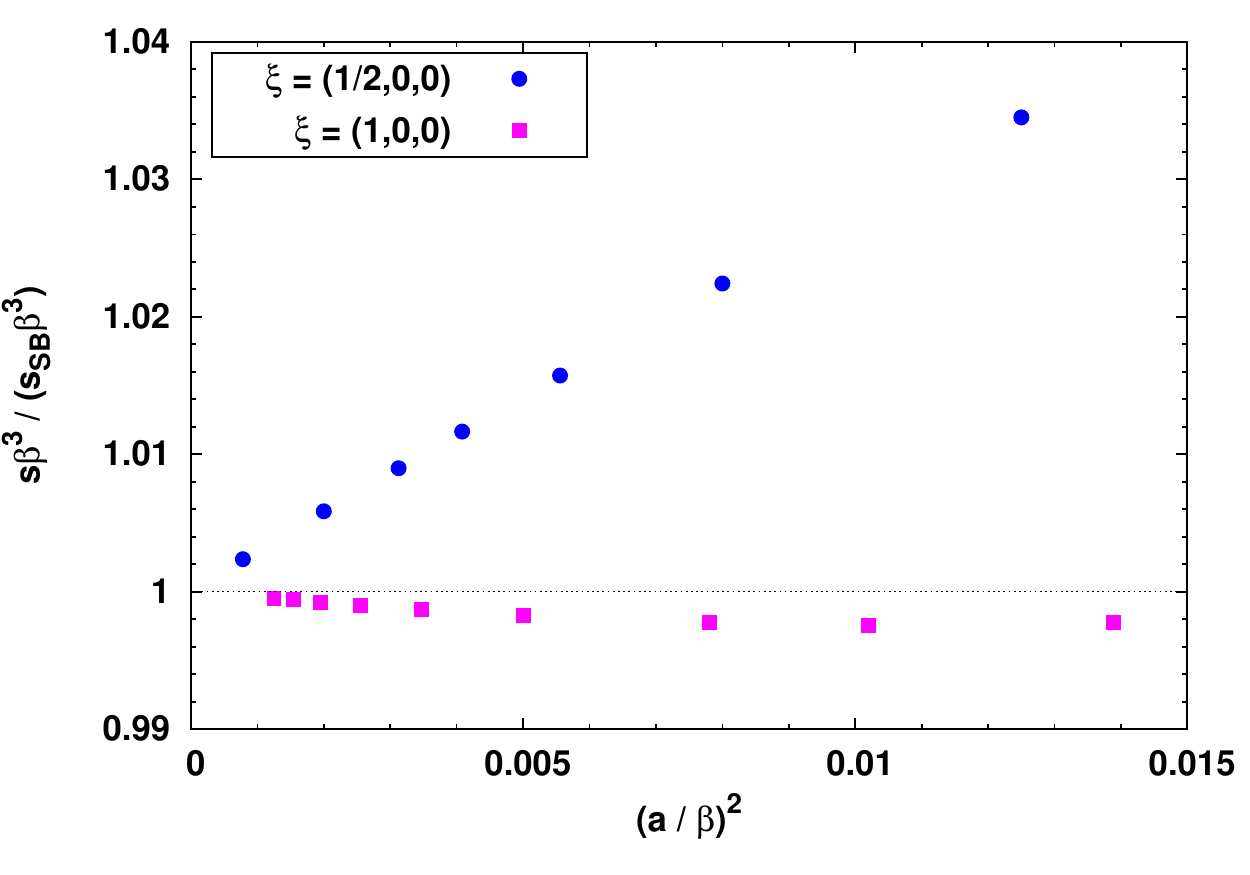}}
\vspace{-0.9cm}
\label{fig:T0k}
\caption{Entropy density at finite lattice spacing for the 
SU($N$) Yang-Mills theory in the non-interacting limit calculated 
via \eq(\ref{eq:rotThMuNu}) and normalized to its continuum 
value $s_{\rm SB}\beta^3=\frac{4\pi^2(N^2-1)}{45}$. The discretization 
used is the Wilson action and the `clover' form of the 
lattice field strength tensor, see appendix \ref{sec:<T0k>}. 
The inverse temperature is $\beta = L_0 \sqrt{1+\vec \xi^2}$,
and $a$ is the lattice spacing.}}
Once the entropy has been computed at various values of $\beta$,
the pressure can be computed by integrating $s$ in the temperature.
The ambiguity left due to the integration constant is consistent
with the fact that $p$ is defined up to an arbitrary 
additive renormalization constant.

The entropy density could also be computed  directly from 
\eq(\ref{eq:stwopt}) without the need for fixing the multiplicative renormalization 
constant. This would require, however, the computation of the two-point 
correlation functions in a large volume. The latter can also be used to access the 
specific heat of the system. From Eqs.~(\ref{eq:stwopt}) 
and (\ref{eq:cvtwopt}), by choosing all $L_k$ and all $\xi_k$ equal, 
for instance, the speed of sound $c_s$ is given by
\be
\frac{1}{c_s^2} = \frac{c_v}{s} = \frac{3}{\vec\xi^2}
\frac{\left\langle {\overline T}_{01}\, {T}_{02}\right\rangle_{\vec\xi,c}+
\vec\xi^2\left\langle {\overline T}_{01}\, {T}_{01}\right\rangle_{\vec\xi,c}}
{\left\langle {\overline T}_{01}\, {T}_{02}\right\rangle_{\vec\xi,c}- 
 \left\langle {\overline T}_{01}\, {T}_{01}\right\rangle_{\vec\xi,c}}\; ,
\ee
where as usual in each correlator all fields  are inserted 
at physical distance. Note that without shifted boundary 
conditions, the specific heat would
require the computation of a four-point function of 
${\overline T}_{0k}$ \cite{Giusti:2010bb,Giusti:2011kt}. Note also 
that all the computational strategies sketched in this 
section use correlation functions of local operators
that require at most an overall renormalization constant. The 
latter can be fixed by WIs in finite volume as described in the 
previous sub-section, and no ultraviolet power-divergent subtractions 
are needed.

\subsection{The integral method at fixed shift}
\FIGURE[t]{
\centerline{\includegraphics[width=12.0 cm,angle=0]{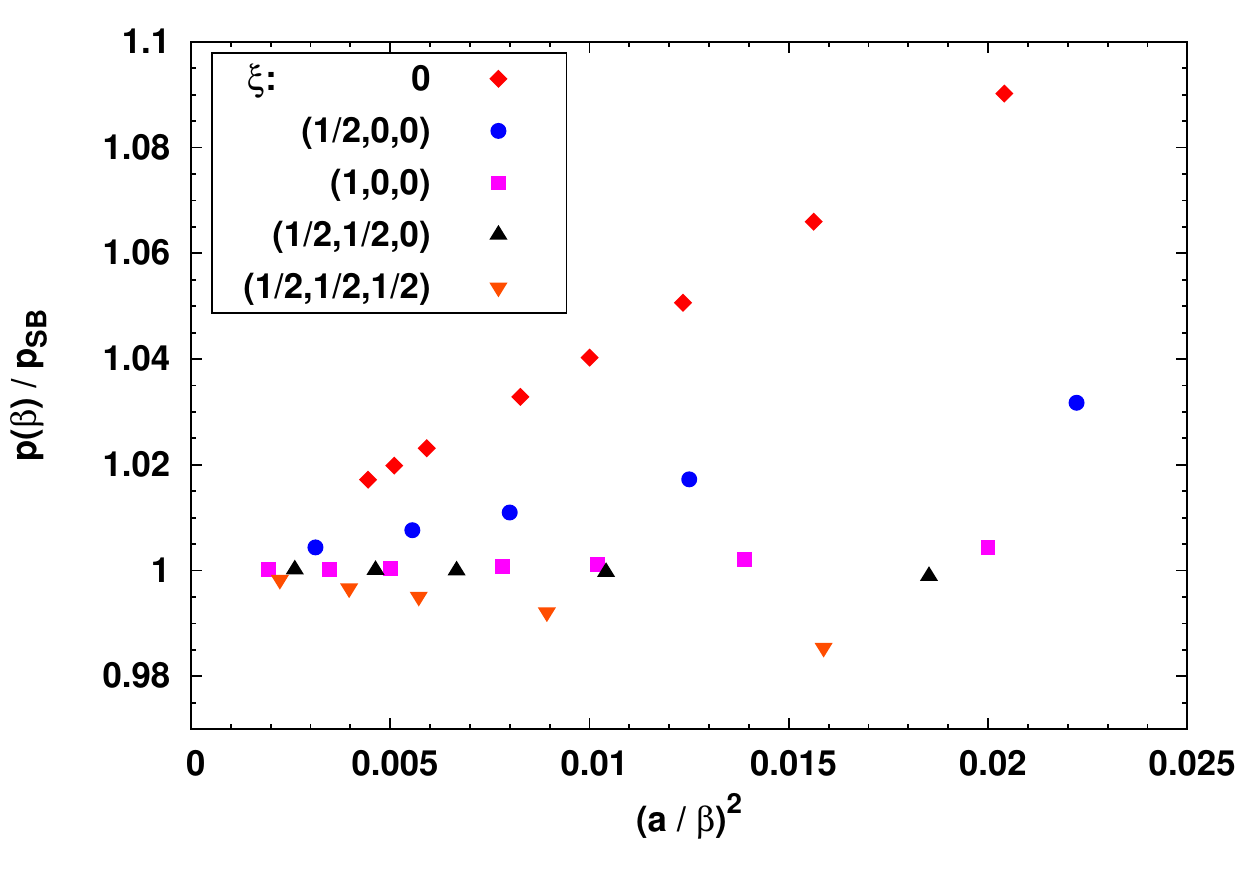}}
\vspace{-1cm}
\label{fig:pressure}
\caption{Pressure at finite lattice spacing for the  
SU($N$) Yang--Mills theory in the non-interacting limit. 
The discretization used is the Wilson action and 
the `clover' form of the lattice field strength tensor. The 
inverse temperature is given by $\beta = L_0 \sqrt{1+\vec \xi^2}$,
and $a$ is the lattice spacing.}}
Calculations of thermodynamic quantities in lattice gauge theories 
(see~\cite{Philipsen:2012nu} for a recent review) usually focus 
on obtaining the pressure $p$. In the thermodynamic limit and 
for a homogeneous system, the pressure is equal to minus the free 
energy, $p=-f$, and all other thermodynamic
potentials can in principle be derived from it by taking derivatives
with respective to the temperature. In the original integral
method proposal~\cite{Engels:1990vr}, the pressure $p$ is computed by
carrying out a line integral of the gradient
of the logarithm of the partition function with respect to the 
bare parameters. The integrand is
expressed as an expectation value of derivatives of the action with
respect to the bare parameters $\vec\alpha=(\alpha_1,\dots,\alpha_n)$. 
Starting the integration from a point
where by convention the free energy vanishes, one obtains
\be\la{eq:IntMeth}
p(L_0) = -\frac{1}{L_0 L^3}\int_{\vec\alpha_{\rm 1}}^{\vec\alpha_{\rm 2}} 
d\vec\alpha\cdot \left\<\nabla_{\vec\alpha} S \right\>_{\vec\alpha}\; ,
\ee
where $S$ is the lattice action and $\<\dots\>_{\vec \alpha}$ is an
expectation value taken at the bare parameter set $\vec \alpha$. 
The integral is done by keeping fixed $L_0/a$, and the temperature
is changed by varying the lattice spacing. The latter is achieved by
varying the bare coupling, but in doing so the other couplings (if any)
must be adjusted if one wants to remain on a line of constant
physics. Moreover a subtraction of the vacuum contribution 
must be made in evaluating the path-integral expectation value of 
$\frac{\partial S}{\partial \alpha_k}$, which is usually done 
at a temperature different from the target temperature, or even at zero 
temperature~\cite{Engels:1990vr,Borsanyi:2012ve}.
Thanks to the integral method, many results have been obtained on the lattice 
for thermal gauge theories~\cite{Philipsen:2012nu}. 
Due to the vacuum subtraction and to the integral on the bare parameters at 
constant physics, it remains difficult, however, to reach large 
temperatures or to apply this method to 
regularizations where the tuning of the bare parameters is technically 
demanding. 

The integral method can also be applied 
in the presence of shifted boundary conditions.  
For a given inverse temperature $\beta=L_0 \sqrt{1+\vec \xi^2}$, 
the continuum limit $\beta/a\to\infty$ of $p\beta^4$ can be
taken at fixed $\vec\xi$. This means that the
angles among the lattice axes and the torus directions are kept fixed. The
discretization effects on $p\beta^4$ in the non-interacting limit of 
the SU($N$) Yang--Mills theory are displayed in \fig\ref{fig:pressure}. 
The plot shows that, in the free theory, they can be drastically reduced 
by using shifted boundary conditions. It remains to be seen whether this 
fact persists in the interacting theory.

\subsection{Temperature scan at fixed lattice spacing}
\FIGURE[t]{
\centerline{\includegraphics[width=12.0 cm,angle=0]{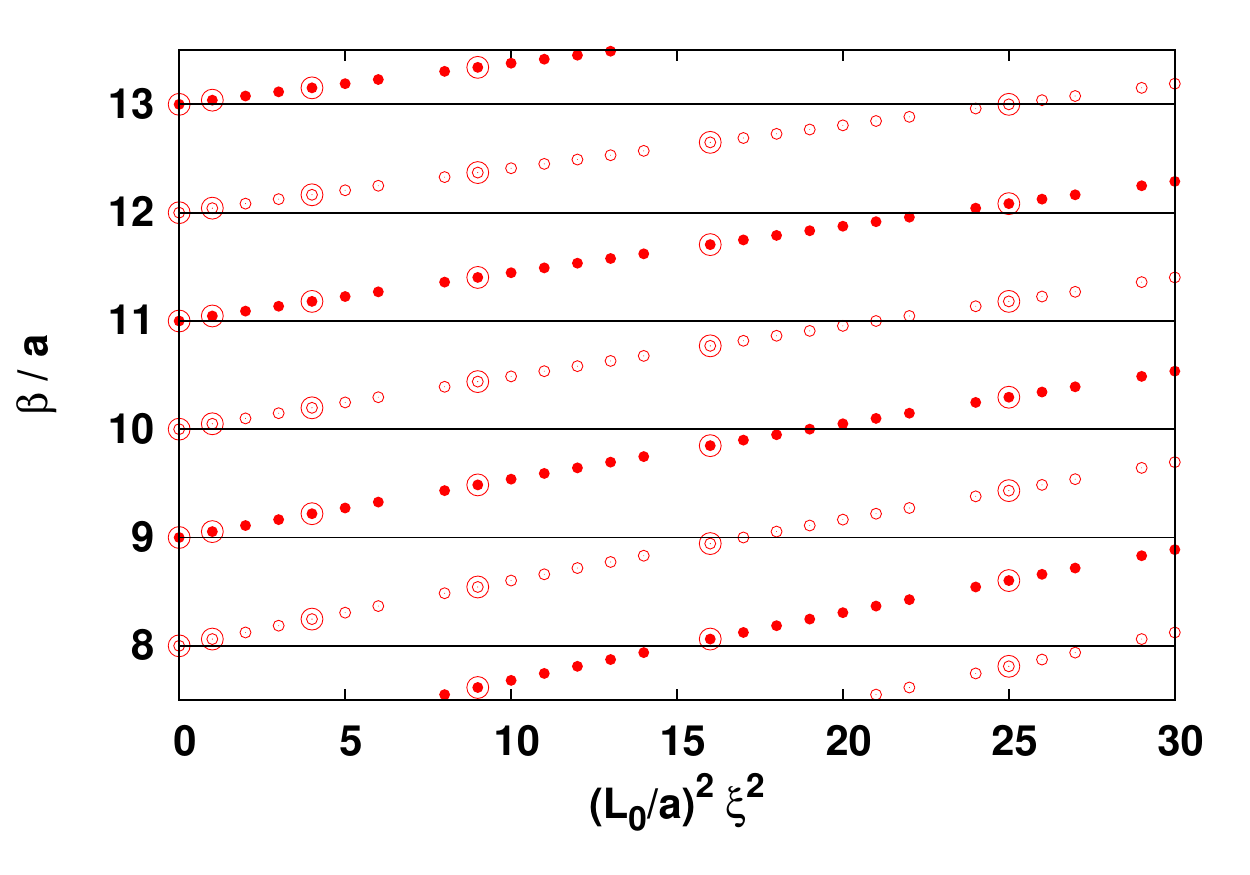}}
\vspace{-1cm}
\label{fig:varytemp}
\caption{Inverse temperature values that become accessible with the use
of shifted boundary conditions at a fixed lattice spacing $a$ and for 
different values of $L_0/a$. The inverse temperatures accessible with a 
shift in a single direction, $\vec \xi = (\xi_1,0,0)$, are marked by 
a double circle.}}
The possibility of varying the temperature by changing
either $L_0/a$ or $\vec\xi$ allows for a fine scan of the
temperature axis at fixed lattice spacing. This is 
illustrated in \fig\ref{fig:varytemp}, where it is also 
compared with the standard procedure of varying $L_0/a$ only.
This fact may turn out to be useful in all those cases where 
the temperature needs to be changed in small steps, e.g.
study of phase transitions etc.

The `T-integral' method~\cite{Umeda:2008bd} is an approach for 
computing thermodynamic quantities related to the integral method 
which is however formulated directly in the continuum. 
Here the pressure is computed as the integral 
\be
p(\beta_2) 
= \frac{\beta^4_1}{\beta_2^4}\; p(\beta_1)  -\frac{1}{\beta_2^4} 
\int_{\beta_1}^{\beta_2} d\beta \beta^3 \, [e(\beta)-3p(\beta)]\; , 
\ee
with the integrand computed by Monte Carlo simulations, and 
the inverse temperature $\beta$ is varied by 
changing $L_0/a$ while keeping 
the bare parameters (bare coupling, quark masses,\dots) of the 
theory fixed. This method has a number of advantages over the method 
based on \eq(\ref{eq:IntMeth}).  The subtraction of the vacuum 
contribution only requires a single zero-temperature simulation, and no 
tuning of bare parameters is required to perform a scan in temperature. 
A significant drawback however is that for a given, realistic 
($L_0/a=8\dots20$) lattice spacing the temperature can only be 
varied in rather coarse steps. The shifted boundary conditions provide a 
way to almost completely eliminate this drawback. Since the integrand 
is a Lorentz-scalar, its expectation value in the
presence of the shifted boundary condition is
equal, up to discretization effects, to its expectation value in the 
unshifted ensemble at inverse
temperature $\beta=L_0\sqrt{1+\vec\xi^2}$. The integrand can thus be 
scanned in much finer steps, and the integral can be computed as 
\be
p(L_0)\!=\! (1+\vec\xi^2)^2 p(L_0\sqrt{1+\vec\xi^2})
+ \frac{1}{2}\!\int_0^{\vec\xi^2}\!\!\!\!\!\! dy (1+y)
\Big[e(L_0\sqrt{1+{\vec\xi'}^{2}})-3p(L_0\sqrt{1+{\vec\xi'}^{2}})
\Big]_{{\vec\xi'}^{2}=y}\!\!\!\!\!\! .
\ee
The shifted boundary conditions and the associated WIs  also suggest
a different implementation of the method. Thanks to \eq(\ref{eq:rotThMuNu}), 
and by remembering that $\beta^2 \frac{\partial}{\partial \beta} p=-s$,
the pressure difference at two temperatures and at a given 
lattice spacing can be computed for instance as 
\be
p(L_0\sqrt{2}) = p\Big(L_0\sqrt{2+\vec\xi_\perp^2}\Big) - 
\frac{Z_{T}}{2} \int_0^{\vec\xi_{\perp}^2} dy\, 
\langle T_{0k} \rangle_{\vec\xi'}\,\Big|_{\vec{\xi'}^2_\perp=y}\; .
\ee
where $\xi'_k=1$ and $\vec\xi'_{\perp}$ stands for 
the two components orthogonal to the $k$-direction.

\section{Conclusions}
Lorentz invariance implies a great degree of redundancy in defining a
relativistic thermal theory in the Euclidean path-integral formalism.
In the thermodynamic limit, the orientation of the compact periodic
direction with respect to the coordinate axes can be chosen at will
and only its length is physically relevant. This redundancy in the
description implies that the total energy and momentum distributions
in the canonical ensemble are related.

For a finite-size system, the lengths of the box dimensions break this
invariance.  The orientation of the Matsubara cycle relative to the
spatial directions does have effects which, however, are exponentially
suppressed. In the limit of large spatial volume the latter are
calculable in terms of the mass and multiplicity of the lightest
screening state(s) of the theory. Being a soft breaking, the
correlation functions of the traceless part of the energy-momentum
tensor still satisfy exact Ward Identities.

When the theory is regularized in the ultraviolet on a hypercubic lattice,
the latter singles out a particular reference frame. The overall
orientation of the periodic cycles of the finite-volume,
finite-temperature system with respect to this preferred coordinate
system affects renormalized observables at the level of lattice
artifacts. As the cutoff is removed, the artifacts are suppressed by a
power of the lattice spacing.

The shifted boundary condition introduced
in~\cite{Giusti:2010bb,Giusti:2011kt} constitute a particularly
interesting instance of the generalized boundary conditions described
in section 3. In the language of the canonical formalism, the energy
eigenstates acquire a phase proportional to their momentum. 
This different but equivalent point of view implies that thermodynamic
potentials can be directly inferred from the response of the partition
function to the shift in the boundary
conditions~\cite{Giusti:2010bb,Giusti:2011kt}, a response which is
also encoded in the expectation value and in the correlators of the
off-diagonal components of the energy-momentum tensor.

The flexibility in the lattice formulation added by the introduction
of a triplet $\vec\xi$ of (renormalized) parameters specifying the
temporal boundary condition has interesting applications. It suggests
new and simpler ways to compute thermodynamic potentials, and the Ward
identities mentioned above can be enforced in a small volume to
determine the renormalization constants of the energy-momentum tensor
components.
The temperature can be changed either by varying $L_0$ in multiples of
the lattice spacing or via the shift parameters $\vec\xi$. This
results in a much finer scan of its value at fixed bare parameters, a
feature that may prove particularly useful in investigations of phase
transitions (see for instance~\cite{Brandt:2012sk}).

\acknowledgments{H.M.\ thanks G.D.\ Moore and S.\ Caron-Huot for illuminating discussions 
during the workshop \emph{Gauge Field Dynamics In and Out of Equilibrium},
March 5 -- April 13, 2012,
held at the Institute for Nuclear Theory in Seattle, WA, USA.
L.G.\ thanks K.\ Rummukainen for discussions. This work was
  partially supported by the MIUR-PRIN contract 20093BMNPR;
by the \emph{Center for Computational Sciences in Mainz};
and by the DFG grant ME 3622/2-1 \emph{Static and dynamic
properties of QCD at finite temperature}.}

\appendix

\section{Derivation of \eq(\ref{eq:k2ncn})\la{sec:s2apdx}}
The \eq(\ref{eq:cn}), which expresses a linear relation
between the $c_1,\dots,c_n$ and the $n$ first derivatives
of the free energy density, is readily inverted
\be\la{eq:ck_inv}
L_0^{n+1}\frac{\partial^n f}{\partial L_0^n} = (-1)^{n+1} n!\sum_{k=1}^n \frac{1}{k!}\, L_0^{k}\, c_k \,,
\qquad n=1,2,\dots 
\ee
On the other hand, after some algebra, it is possible to show that 
\be\la{eq:x.d.dx}
\left(\frac{1}{L_0} \frac{\partial}{\partial L_0}\right)^n f =
\frac{1}{L_0^{2n}} \sum_{k=1}^n \frac{(-1)^{n-k}}{2^{n-k}}
\frac{(2n-k-1)!}{(k-1)! (n-k)!}\, L_0^k\, 
\frac{\partial^k f}{\partial L_0^k}\,, \qquad n=1,2,\dots 
\ee
Using first \eq(\ref{eq:x.d.dx}) and then (\ref{eq:ck_inv}), the
derivatives of the free energy in expression (\ref{eq:fnd1}) can be
replaced by the $c_n$. One then arrives at the desired relation between 
the cumulants of the momentum and the energy operator, 
\eq(\ref{eq:k2ncn}).

\section{Momentum-space analysis of GPBCs\label{app:MOM}}
Since translational invariance is left unbroken by the boundary conditions,
we can expand the fields in Fourier modes. The set of momenta 
compatible with the boundary conditions is\footnote{In crystallographic
terminology this is the \emph{reciprocal lattice}.}
\be\la{eq:GammaV}
\Gamma_V = \Big\{ p \in \mathbb{R}^d \;|\; p_\mu = 2\pi (V^{-1})_{\nu\mu} n_\nu, ~
n_\nu\in \mathbb{Z}\Big\}
\ee
and the plane wave expansion reads
\be\la{eq:plwe}
\phi_\sigma(x) = \sum_{p\in\Gamma_V} \tilde\phi_\sigma(p) \; e^{ip\cdot x}
 = \phi_\sigma(x+ V m)\,,\quad (Vm)_\mu = V_{\mu\nu} m_\nu\,, \quad m_\nu\in\mathbb{Z}\, .
\ee
Clearly, we have the property
\be\la{eq:GammaEquiv}
\Gamma_{V} = \Gamma_{V\!M},\qquad M\in {\rm SL}(d,\mathbb{Z})\; .
\ee
It is instructive to see
how the equivalence (\ref{eq:equiv}) shows up in momentum space.
First, consider a term $S^\infty_n$ in the action of the 
infinite-volume theory
\be 
S^\infty_n([\phi])\!=\!\!\int\left(\prod_{i=1}^n\frac{d^dp^i}{(2\pi)^d}\right)
c_n^{\sigma_1\dots\sigma_n}\left(p^1,\dots,p^n\right)\,
\tilde\phi_{\sigma_1}(p^1)\cdots \tilde\phi_{\sigma_n}(p^n)\,
(2\pi)^d\; \delta^{(d)}\Big(\sum_{i=1}^n p^i\Big),\!\!   
\ee 
where as usual the delta function enforces momentum conservation 
in all $d$ directions. Since the fields are irreducible 
representations, Lorentz symmetry is encoded in the property
\ba\la{eq:LorentzProp}
&& c_n^{\sigma_1\dots\sigma_n}\left( p^1,\dots, p^n\right)
\tilde\phi_{\sigma_1}^\Lambda(p^1)\cdots \tilde\phi_{\sigma_n}^\Lambda(p^n)
\\ && \qquad \qquad \qquad \qquad 
= c_n^{\sigma_1\dots\sigma_n}\Big(\Lambda^{-1}p^1,\dots,\Lambda^{-1}p^n\Big) 
\tilde\phi_{\sigma_1}(\Lambda^{-1}p^1)\cdots \tilde\phi_{\sigma_n}(\Lambda^{-1}p^n),\qquad 
\nonumber
\ea
i.e.\ the action density at momenta $(p_1,\dots,p_n)$ of the rotated field
\be\la{eq:phiLbda}
\tilde\phi_\sigma^\Lambda(p) \equiv 
U(\Lambda)_{\sigma\sigma'}\; \tilde\phi_{\sigma'}(\Lambda^{-1} p)
\ee
is the same as the action density of the original field at momenta
$(\Lambda^{-1}p_1,\dots,\Lambda^{-1}p_n)$.  This property guarantees
in particular that the infinite-volume action of a rotated field is
equal to the action of the original field.

In the finite-volume theory, the same form of the action holds, but
the integral over momenta is replaced by a sum over the set $\Gamma_V$
of momenta compatible with the periodicity of the field.  We can
write this contribution to the action as
\be
S_n(V;[\phi]) = \frac{1}{V_d^{n-1}}\sum_{p^i\in \Gamma_V}
c_n^{\sigma_1\dots\sigma_n}\left(p^1,\dots,p^n\right)
~\delta_{\sum_{i=1}^n p^i}\; 
\tilde\phi_{\sigma_1}(p^1)\cdots \tilde\phi_{\sigma_n}(p^n)\; ,
\ee
where $V_d$ is the volume of the primitive cell.
Clearly \eq(\ref{eq:GammaEquiv}) implies that 
the action of the two systems parameterized by $V$ and $VM$ are equal
\emph{for the same field}, 
\be\la{eq:SequivM}
S_n(V;[\phi]) = S_n(V M;[\phi])\; .
\ee
Second, we can also write
\be
S_n(V;[\phi]) = \!\frac{1}{V_d^{n-1}}\!\!\sum_{p^i\in\Gamma_{\Lambda V}}
c_n^{\sigma_1..\sigma_n}\left(\Lambda^{-1} p^1,.. ,\Lambda^{-1} p^n\right)
~\delta_{\sum_{i=1}^n p^i}\;
\tilde\phi_{\sigma_1}(\Lambda^{-1} p^1)..\tilde\phi_{\sigma_n}(\Lambda^{-1} p^n)\, ,
\ee
and by using \eq(\ref{eq:LorentzProp}) one finds 
\be\la{eq:SequivLbda}
S_n(V;[\phi]) = S_n(\Lambda V;[\phi^\Lambda])\; . 
\ee
One immediate implication of \eq(\ref{eq:SequivM}) and (\ref{eq:SequivLbda}) 
is that the partition functions of the same field theory with equivalent
sets of boundary conditions $V$ and $W=\Lambda V M $ are equal
\be\la{eq:ZVeqZW}
V\sim W \qquad \Rightarrow \qquad Z(V) = Z(W)\; .
\ee

\subsection{An alternative representation of a field theory with GPBCs} 
We return briefly to the plane wave expansion of the field, \eq(\ref{eq:plwe}),
to mention an alternative representation of a field theory with GPBCs.
The rotation matrix $\Lambda$ can always be chosen such that $V$ is 
triangular. By denoting $V_{\mu\mu} = L_\mu$, we can write $V =
(R^\intercal)^{-1} D$, where $R$ is triangular and all its
diagonal elements are unity, and $D={\rm diag}(L_0,\dots,L_{d-1})$.
Consider then the field transformation
\be\la{eq:hatphidef}
\hat\phi_\sigma(x)\equiv \phi_\sigma({R^{-1}}^\intercal x)\; . 
\ee
By expanding $\hat\phi(x)$ in Fourier modes 
\ba\la{eq:hatphi}
\hat\phi_\sigma(x) &=& \sum_{p=2\pi {D^{-1}} n} {\tilde{\hat\phi}}_\sigma(p)\; e^{ip\cdot x},
\ea
in momentum space \eq(\ref{eq:hatphidef}) becomes
\be
\tilde{\hat\phi}_\sigma(p) = \tilde{\phi}_\sigma(R p)\; .
\ee
From both \eq(\ref{eq:hatphidef}) and \eq(\ref{eq:hatphi}), it is clear that $\hat\phi$
fulfills ordinary periodic boundary conditions on a primitive cell
with $d$ orthogonal sides of lengths $(L_0,\dots,L_{d-1})$.
The effect of the non-orthogonality of the original primitive vectors is 
absorbed into the action for $\hat\phi$,
\be
\!\!\!\!\hat S_n(V;[\hat\phi]) \equiv S_n(V;[\phi]) =\! 
\frac{1}{V_d^{n-1}}\!
\sum_{p^i\in\Gamma_D} c_n^{\sigma_1..\sigma_n}(R p^1,\dots,R p^n)
~\delta_{\sum_{i=1}^n p^i}\; \tilde{\hat\phi}_{\sigma_1}(p^1)..
\tilde{\hat\phi}_{\sigma_n}(p^n).
\ee
The kinetic term of a scalar field theory, for instance,  
\[ 
c_2(R p^1,R p^2) = - {\txts\frac{1}{2}}\;{p^1}^\intercal \,(R^\intercal R) \, p^2
\]
is a positive-definite quadratic form in this formulation.

\section{Free energy of a non-interacting bosonic theory with 
shifted boundaries\la{sec:apdxfree}}
In this appendix we compute the free-energy of a non-interacting 
bosonic field theory in a finite volume. This serve to check  
\eq(\ref{eq:finalFV}) in the free theory explicitly, and it 
also shows that the latter predicts correctly the leading finite-size 
effects for a generic thermal theory if the mass and the multiplicity 
are fixed to those of the lightest screening state(s).

For a generic set of GPBCs, the SO(4) symmetry allows one to 
cast the primitive matrix ${\cal V}$ in the form
\be
{\cal V} = 
\left(\begin{array}{c@{~~~}c@{~~~}c@{~~~}c}
L_0    & 0 & 0 & 0 \\
z_1       &   &   &   \\
z_3 &   & \overline{\Huge {\cal V}}  &   \\
z_3       &   &   & \\
\end{array} \right)\; , 
\ee
see \eq(\ref{eq:Lambda}). The matrix ${\overline {\cal V}}$ 
specifies the spatial periodic directions of the system, and the 
associated three-dimensional reciprocal lattice can be extracted 
from \eq(\ref{eq:GammaV}). It is easy to prove that
\ba
f({\cal V})-\lim_{L_0\to\infty} f({\cal V})
&=& \frac{1}{L_0 \det\overline {\cal V}} \sum_{\bp}
\log(1-e^{-L_0\omega_{\bp}+i\bp\cdot\bz})\nonumber\\ 
&=& \frac{2}{L_0\det\overline {\cal V}} \frac{\partial}{\partial L_0} 
\sum_{n\geq 1} \frac{1}{n^2} \sum_{\bp} 
\,\frac{e^{n(-L_0\omega_{\bp}+i\bp\cdot\bz)}}{2\omega_{\bp}}\; ,
\ea
where $\omega_{\vec p}=\sqrt{\vec p^2 + M^2}$. Thanks to the Poisson 
summation formula
\be
\frac{1}{\det\overline {\cal V}}\sum_{\bp} \,
\frac{e^{n(-L_0\omega_{\bp}+i\bp\cdot\bz})}{2\omega_{\bp}} = 
\sum_{\vec k\in\mathbb{Z}^3} \Delta^4(r,M^2)\Big|_{r=\sqrt{n^2L_0^2+(n \vec z+\overline 
{\cal V}\vec k)^2}}\; , 
\ee
where in $d$-dimensions
\be
\Delta^d(|x|,M^2) \equiv \int \frac{d^d p}{(2\pi)^d} \frac{e^{ipx}}{p^2+M^2}\; .
\ee
We can thus write
\be
f({\cal V})-\lim_{L_0\to\infty} f({\cal V}) =  \sum_{k_0\neq 0}
\sum_{\vec k}
 \left[\frac{1}{r}\frac{\partial}{\partial r} \Delta^4(r,M^2)
\right]_{r=|{\cal V} k|}\;,\qquad
k^\intercal=(k_0,\vec k)\in\mathbb{Z}^4\; .
\ee
By repeating the argument in all $k$-directions, i.e. 
by sending successively $V_{kk}$ to infinity, we arrive at the 
master equation
\be\la{eq:main_free}
f({\cal V}) - f_\infty = 
\sum_{k\neq0}\left[\frac{1}{r}\frac{\partial}{\partial r}
\Delta^4(r,M^2)
\right]_{r=|{\cal V} k|}\; .
\ee
where $f_\infty$ is the free energy of the system on 
$\mathbb{R}^4$, i.e.\ in infinite volume.
Since \eq(\ref{eq:main_free}) is expressed in terms of the norm of all the position 
vectors equivalent by periodicity to the origin, its form is invariant 
within an equivalence class of primitive matrices. It therefore 
holds for any ${\cal V}\in GL(4,\mathbb{R})$. \\[-0.325cm]

As an application of \eq(\ref{eq:main_free}), we consider the case where 
${\cal V}$ is equal to $V_{\rm sbc}$ defined in \eq(\ref{eq:Vsbc}), i.e.  
\be\la{eq:J}
f(V_{\rm sbc}) - \lim_{L_1,L_2,L_3\to\infty} f(V_{\rm sbc}) =  {\cal J} 
\equiv  \sum_{n\in\mathbb{Z}} \sum_{\vec m\neq 0} 
\left[
\frac{1}{r}\frac{\partial}{\partial r} \Delta^4(r,M^2) 
\right]_{r=\sqrt{n^2L_0^2+(nL_0\vec\xi + \vec\mu)^2}}
\ee
where $\vec\mu=(m_1 L_1,m_2 L_2,m_3 L_3)$. Expression (\ref{eq:J}) involves 
4d propagators, while \eq(\ref{eq:resultEmr}) contains 3d propagators.  
Using again the Poisson formula, one obtains
\be
{\cal J} = 
 \sum_{m_0\in\mathbb{Z}} \sum_{\vec m\neq 0} \int_{-\infty}^\infty d\eta
\; e^{i2\pi m_0 \eta} \;
\left[\frac{1}{r}\frac{\partial}{\partial r} \Delta^4(r,M^2) \right]_{r=\sqrt{\eta^2L_0^2+(\eta L_0\vec\xi
+\vec\mu)^2}}\; , 
\ee
where the argument of the propagator can be rewritten as
\be
r^2 = \eta^2L_0^2+(\eta L_0\vec\xi+\vec\mu)^2 = 
\frac{1}{\gamma^2} \left(\eta L_0+ \gamma^2 \vec\xi\cdot\vec\mu\right)^2
+  (Q\vec\mu, Q\vec\mu)\; 
\ee
with $Q_{ij}$ being defined below \eq(\ref{eq:resultEmr}). By setting
$x_0 =(\eta L_0/\gamma + \gamma\,\vec\xi\cdot\vec\mu)$,
and by using the `dimensional reduction' relation
between the four- and three-dimensional propagators
\be
\int_{-\infty}^\infty \frac{dx_0}{x_0}\; e^{-i\omega x_0} \;
\frac{\partial}{\partial x_0} 
\Delta^4(|x|,M^2) 
= \left[\frac{1}{r}\frac{\partial}{\partial r} 
\Delta^3(r,\omega^2+M^2) \right]_{r=|\vec x|}\; ,
\ee
the following result emerges
\be
{\cal J} = \frac{\gamma}{L_0}\sum_{m_0} \sum_{\vec m \neq 0} 
e^{-i2\pi m_0 \gamma^2(\vec\xi\cdot\vec\mu)/L_0} 
\left[\frac{1}{r}\frac{\partial}{\partial r} 
\Delta^3\left(r,M^2+(2\pi m_0\gamma/L_0)^2\right)
\right]_{r=|Q\vec\mu|}\; .
\ee
For large spatial box dimensions $L_k$, we can drop all terms but $m_0=0$
and obtain 
\be
{\cal J} = \frac{\gamma}{L_0}\sum_{\vec\mu\neq 0}\left[
\frac{1}{r}\frac{\partial}{\partial r} \Delta^3(r,M^2)\right]_{r=|Q\vec\mu|}+\dots\; , \qquad
\Delta^3(r,M^2) = \frac{1}{4\pi r} e^{-M r}\; .
\ee
Taking into account that the relevant terms in $\vec\mu$ that give the
leading correction in $L_1$ are now $\mu_1=\pm L_1$, $\mu_2=\mu_3=0$, we
recover exactly \eq(\ref{eq:resultEmr}) and therefore 
\eq(\ref{eq:finalFV}), if we set the mass and the multiplicity in the 
free-theory equal to those of the lightest screening state in the 
interacting theory.

\section{Free bosonic theory on the lattice with shifted boundary 
conditions\la{sec:<T0k>}}
On a finite-volume lattice specified by the primitive matrix 
$V_{\rm sbc}$ in \eq(\ref{eq:Vsbc}), the bosonic propagator in position 
space reads\footnote{The lattice spacing is set to $a=1$ in this 
appendix.}
\be\label{eq:latprop}
\Delta_{\rm L}^4(x,M^2) = \frac{1}{L_0 L_1 L_2 L_3}
\sum_{\ell=0}^{L_0-1} \sum_{\vec p\in {BZ}} 
\frac{e^{i(\frac{2\pi\ell}{L_0} - \vec p\cdot \vec \xi)x_0 
+ i\vec p\cdot \vec x}}
{ 4 \sin^2(\frac{\pi\ell}{L_0} - \frac{\vec p\cdot \vec \xi}{2})
  + M^2 + 4 \sum_{k=1}^3 \sin^2(\frac{p_k}{2})}\; ,
\ee
where $BZ$ stands for the Brillouin zone. We are thus interested 
in the finite sum
\be
\Sigma(x_0)=\frac{1}{L_0}\sum_{\ell=0}^{L_0-1} 
\frac{e^{ix_0(2\pi \ell/L_0 -  \phi)}}
{\omega^2 + 4\sin^2(\frac{\pi\ell}{L_0} - \frac{\phi}{2})}\; ,
\ee
and for each value of $\vec p$ we will set 
\be\label{eq:subs}
\phi = \vec p\cdot\vec \xi\; , \qquad
\omega^2=M^2 + 4 \sum_{k=1}^3 \sin^2(\frac{p_k}{2})
\ee 
at the end of the calculation. To this end we generalize a well known 
contour integral calculation, see for instance 
Ref.~\cite{Elze:1988zs}. The first observation is that 
\be\la{eq:SL}
\Sigma(x_0) = \frac{1}{L_0}\sum_{\ell=0}^{L_0-1}
g(e^{i\pi\ell/L_0-i\phi/2},x_0)\; ,\qquad g(z,x_0) = 
\frac{z^{2x_0}}{\omega^2- (z - z^{-1})^2}\; ,
\ee
and using the fact that $g(z,x_0)=g(-z,x_0)$ we have
\be\la{eq:S2L}
\Sigma(x_0) = \frac{1}{2L_0}\sum_{\ell=0}^{2L_0-1}
g(e^{i\pi\ell/L_0-i\phi/2},x_0)\; .
\ee
The poles of $g(z,x_0)$ in the variable $z$ are on the real axis at 
\be
\bar z_{1,\dots,4} =  \pm \frac{\omega}{2} \pm 
\sqrt{(\frac{\omega}{2})^2 + 1}\; .
\ee
Consider the integral
\be\la{eq:Iall}
I_{\rm all}=\int_{\Gamma_{\rm all}} \frac{dz}{z} \;\frac{g(z,x_0)}{e^{iL_0\phi}z^{2L_0}-1}\; ,
\ee
where the contour $\Gamma_{\rm all}$ contains all the singularities
of the integrand. The latter are all simple poles located at
\be
\bar z_1,~ \bar z_2,~\bar z_3,~\bar z_4\; ;\qquad 
\hat z_\ell = e^{i\pi\ell/L_0-i\phi/2},\qquad \ell=0,1,\dots(2L_0-1)\; .
\ee
The integral $I_{\rm all}$ vanishes since the integrand 
falls off as $|z|^{-5}$ when $|z|\rightarrow\infty$.
As a consequence the sum of all residues is null, and 
the sum of the residues at $\bar z_i$ equals minus the sum of
the residues at $\hat z_\ell$. Since
\be
e^{iL_0\phi}\!z^{2L_0}\!\!-1\! =\! (z_\phi-e^{i\pi\ell/L_0})\!
\Big[ z_\phi^{2L_0-1}\! + e^{i\pi\ell/L_0}z_\phi^{2L_0-2}\!+..+
e^{i\pi\ell(2L_0-2)/L_0} z_\phi + e^{i\pi(2L_0-1)/L_0}\Big]
\qquad\!\!\!\!\!\!\!\!\!\!\!\!\!\!\!
\ee
with $ z_\phi \equiv z e^{i\phi/2}$, the residue of the 
integrand in \eq(\ref{eq:Iall}) at $\hat z_\ell$ is 
\be
\frac{1}{2L_0} g(e^{i\pi\ell/L_0-i\phi/2},x_0)\; .
\ee
Comparing with \eq(\ref{eq:S2L}) we have 
\be
\Sigma(x_0) = \frac{1}{2\pi i} \int_{\hat\Gamma} \frac{dz}{z} 
\frac{g(z,x_0)}{e^{iL_0\phi}z^{2L_0}-1} = 
- \frac{1}{2\pi i} \int_{\bar\Gamma} \frac{dz}{z} 
\frac{g(z,x_0)}{e^{iL_0\phi}z^{2L_0}-1}\; ,
\ee
where the contour $\hat\Gamma$ encircles the poles 
$\hat z_\ell$ but not the poles $\bar z_i$, while for $\bar\Gamma$
it is the other way around. Since
\be
{\rm Res}\left( \frac{1}{z} \frac{1}{\omega^2-(z-z^{-1})^2} \right)_{z=\bar z_i} = 
 \frac{1}{2(1/\bar z_i^{2} - \bar z_i^2)}\; ,
\ee
then
\be
\Sigma(x_0)= -\frac{1}{2} \sum_{i=1}^4 \frac{\bar z_i^{2x_0}}{e^{iL_0\phi}\bar z_i^{2L_0}-1}
\; \frac{1}{1/\bar z_i^{2} - \bar z_i^2}\; .
\ee
By setting $\omega = 2\sinh(\hat\omega/2)$, it then follows 
that $\bar z_i^2 = e^{\pm\hat\omega}$ and 
\ba\label{eq:sigmax0}
\Sigma(x_0)&=& \frac{1}{2\sinh\hat\omega}\left[  
\frac{e^{\hat\omega x_0}}{e^{iL_0\phi+L_0\hat\omega}-1} - 
\frac{e^{-\hat\omega x_0}}{e^{iL_0\phi-L_0\hat\omega}-1} \right].
\ea
The real and imaginary parts read
\ba\la{eq:reSig}
\re\Sigma(x_0)&=& 
\frac{\sinh(L_0\hat\omega/2)\cosh[\hat\omega(L_0/2-x_0)] - \sin^2(L_0\phi/2) \sinh(\hat\omega x_0)}
{\sinh\hat\omega\; \left(\cosh(L_0\hat\omega) - 
\cos(L_0\phi)\right)}\;,\\[0.25cm]
\im\Sigma(x_0)&=& \frac{-\sin(L_0\phi)\,\sinh(\hat\omega x_0)}
{2\sinh(\hat\omega)\,\left(\cosh(L_0\hat\omega) - \cos(L_0\phi)\right)}
\; .
\la{eq:imSig}
\ea
For $\phi=0$, corresponding to periodic boundary conditions, one recovers the known (real) result
\be
\Sigma(x_0) = \frac{1}{2\sinh\hat\omega}\; \frac{\cosh[\hat\omega(L_0/2-x_0)]}{\sinh(L_0\hat\omega/2)}\; .
\ee
Finally the propagator is obtained by 
inserting \eq(\ref{eq:sigmax0}) in \eq(\ref{eq:latprop}) for 
each value of $\vec p$ after having made the substitutions 
in \eq(\ref{eq:subs}).

\subsection{Expectation value of $T_{0k}$ for the SU($N$) gauge theory}
We are interested in the expectation value of the momentum density 
operator in the non-interacting limit of the SU($N$) gauge 
theory in presence of shifted boundary conditions. 
We discretize $T_{01}$ using the `clover' discretization of the 
field strength tensor as described in~\cite{Meyer:2008sn,Meyer:2009vj}.  
Using the perturbative expansion and taking the infinite volume 
limit we obtain~\cite{Meyer:2009vj}
\ba
\frac{1}{2 (N^2-1)}\<T_{01}\> &=& 
\frac{1}{L_0} \sum_{\ell=0}^{L_0-1} \int_{BZ} \frac{d^3\vec p}{(2\pi)^3}
\; \frac{\sin(2\pi\ell/L_0 - \vec p\cdot\vec\xi) \cos^2(p_2/2)\sin(p_1)}
{4\sin^2(\frac{\pi\ell}{L_0} - \frac{\vec p\cdot \vec\xi}{2}) 
+ 4\sum_{k=1}^3 \sin^2(p_k/2)} \nonumber\\[0.25cm] 
&=&  \int_{BZ} \frac{d^3\vec p}{(2\pi)^3}  \cos^2(p_2/2)\sin(p_1)\; \im\Sigma(1)\; ,
\la{eq:T01vev}
\ea
where in the latter equation for $\Sigma(1)$ we use 
\eq(\ref{eq:imSig}) with $\omega^2=4\sum_{k=1}^3 \sin^2(p_k/2)$ 
and $\phi = \vec p\cdot\vec\xi$.  
The three-dimensional integral in \eq(\ref{eq:T01vev}) can be evaluated 
numerically leading to \fig\ref{fig:T0k}.

\bibliographystyle{JHEP}
\bibliography{/home/meyerh/CTPHOPPER/ctphopper-home/BIBLIO/viscobib.bib}

\end{document}